\documentclass[%
 reprint,
 amsmath,amssymb,
floatfix,
]{revtex4-1}

\usepackage{graphicx}% Include figure files
\usepackage{dcolumn}% Align table columns on decimal point
\usepackage{bm}% bold math
\usepackage{color}
\usepackage{comment}
\usepackage{hyperref}% add hypertext capabilities
\newcommand{\ve}[1]{\vec{#1}}
\newcommand{\vx}{\ve{x}}

\newcommand{\veps}{\ve{\epsilon}}
\newcommand{\vw}{\ve{w}}

\newcommand{\vv}{\ve{v}}

\newcommand{\Ol}{\mathcal{O}}

\newcommand{\degree}{\ensuremath{^\circ}}

\newcommand{\op}{\text{SYRTE}}
\newcommand{\npl}{\text{NPL}}
\newcommand{\ptb}{\text{PTB}}

\newcommand{\nplop}{\text{NPL-SYRTE}}
\newcommand{\ptbop}{\text{PTB-SYRTE}}

\newcommand{\ySr}{{\bar y}^{\text{Sr2}}_{\ptbop}}

\begin{document}

%\begin{comment}
\title{Test of special relativity using a fiber network of optical clocks}% Force line breaks with \\

\author{P. Delva$^1$}
\email{Pacome.Delva@obspm.fr}
%\begin{comment}
\author{J. Lodewyck$^1$}
\author{S. Bilicki$^1$}
\author{E. Bookjans$^1$}
\author{G. Vallet$^1$}
\author{R. Le Targat$^1$}
\author{P.-E. Pottie$^1$}
\author{C. Guerlin$^{2,1}$}
\author{F. Meynadier$^1$}
\author{C. Le Poncin-Lafitte$^1$}
\author{O. Lopez$^3$}
\author{A. Amy-Klein$^3$}
\author{W.-K. Lee$^{1,4}$}
\author{N. Quintin$^3$}
\author{C. Lisdat$^5$}
\author{A. Al-Masoudi$^5$}
\author{S. D\"orscher$^5$}
\author{C. Grebing$^5$}
\author{G. Grosche$^5$}
\author{A. Kuhl$^5$}
\author{S. Raupach$^5$}
\author{U. Sterr$^5$}
\author{I. R. Hill$^6$}
\author{R. Hobson$^6$}
\author{W. Bowden$^6$}
\author{J. Kronj\"{a}ger$^6$}
\author{G. Marra$^6$}
\author{A. Rolland$^6$}
\author{F. N. Baynes$^6$}
\author{H. S. Margolis$^6$}
\author{P. Gill$^6$}

\affiliation{%
$^1$SYRTE, Observatoire de Paris, PSL Research University, CNRS, Sorbonne Universit\'es, UPMC Univ. Paris 06, LNE, 61 avenue de l'Observatoire 75014 Paris, France
}%
\affiliation{%
$^2$Laboratoire Kastler Brossel, ENS-PSL Research University, CNRS, UPMC-Sorbonne Universit\'es, Coll\`ege de France, 75005 Paris, France
}
\affiliation{%
$^3$Laboratoire de Physique des Lasers, Universit\'e Paris 13, Sorbonne Paris Cit\'e, CNRS, 99 Avenue Jean-Baptiste Cl\'ement, 93430 Villetaneuse, France
}
\affiliation{$^4$Korea Research Institute of Standards and Science, Daejeon 34113, South Korea}
\affiliation{$^5$Physikalisch-Technische Bundesanstalt, Bundesallee 100, 38116 Braunschweig, Germany}

\affiliation{
$^6$National Physical Laboratory, Hampton Road, Teddington, TW11 0LW, UK
}
%\end{comment}
\begin{abstract}
Phase compensated optical fiber links enable high accuracy atomic clocks separated by thousands of kilometers to be compared with unprecedented statistical resolution. By searching for a daily variation of the frequency difference between four strontium optical lattice clocks in different locations throughout Europe connected by such links, we improve upon previous tests of time dilation predicted by special relativity. We obtain a constraint on the Robertson--Mansouri--Sexl parameter $|\alpha|\lesssim 1.1 \times10^{-8}$ quantifying a violation of time dilation, thus improving by a factor of around two the best known constraint obtained with Ives--Stilwell type experiments, and by two orders of magnitude the best constraint obtained by comparing atomic clocks. This work is the first of a new generation of tests of fundamental physics using optical clocks and fiber links. As clocks improve, and as fiber links are routinely operated, we expect that the tests initiated in this paper will improve by orders of magnitude in the near future. 
\end{abstract}

\maketitle
%\end{comment}

Special Relativity (SR), one of the cornerstones of modern physics, assumes that Lorentz Invariance (LI) is a fundamental symmetry of nature. The search for a violation of LI is motivated by two factors: (i) theoretical suggestions that LI may not be an exact symmetry at all energies and (ii) the tremendous advances in the precision of experimental tests. Indeed, a strong violation of LI at the Planck scale is likely to yield a small amount of violation at low energy, which could be measured with precise experiments~\cite{Mattingly2005}. 

Optical clocks are now the most precise measurement devices. They reach systematic uncertainties of a few $10^{-18}$, which can be resolved after a mere few hours of measurement with optical lattice clocks based on trapped neutral atoms~\cite{RevModPhys.87.637,ushijima2015cryogenic,nicholson2015systematic,alm15,sch16c}. Thanks to these unparalleled performances, comparing the resonance frequencies of optical clocks has led to new tests of fundamental physics, such as bounding the time variation of fundamental constants~\cite{PhysRevLett.113.210802,Godun2014}.

In this paper, we perform a test of SR using a network of distant optical lattice clocks located in France, Germany and the UK. By exploiting the difference between the velocities of each clock in the inertial geocentric frame, due to their different positions on the surface of the Earth, we are able to improve upon previous tests of time dilation. The connection between these clocks, achieved with phase-compensated optical fibers, allows for an unprecedented level of statistical resolution for the comparison of remote atomic clocks~\cite{lisdat2015clock}, making such a test possible for the first time. 

LI violations are predicted by several theoretical frameworks, categorized as kinematical and dynamical frameworks (see~\cite{Mattingly2005} for a review). In this paper we use the Robertson--Mansouri--Sexl (RMS) kinematical framework~\cite{robertson49, mansouri77a, mansouri77b, mansouri77c} which contains only three parameters. It assumes the existence of a preferred frame $\Sigma$ where light propagates rectilinearly and isotropically in free space with constant speed $c$. The ordinary Lorentz transformations from $\Sigma$ to the observer frame $S$ with relative velocity $\vw$ are generalized to allow for violations of SR:
\begin{eqnarray}
\label{eq:TR1}
T &=& a^{-1} (t - c^{-1} \veps \cdot \vx)\\
\label{eq:TR2} \ve{X} &=& d^{-1} \vx - (d^{-1} - b^{-1}) (\vw \cdot \vx) \vw /
w^2 + \vw T \ ,
\end{eqnarray}
where $a$, $b$ and $d$ are functions of $w^2$, and $\veps$ is a $w$-dependent
vector specifying the clock synchronization procedure in $S$. In the low-velocity limit:
\begin{equation}
	a(\vw) = 1 + c^{-2} (\alpha-1/2) w^2 + \Ol(c^{-4} w^4) \ , \label{eq:alpha}
\end{equation}
where $\alpha$ is an arbitrary parameter quantifying the LI violation, the value of which is zero in SR.

First-order tests in $\ve{w}/c$ are based on the comparison of clocks~\cite{mansouri77b}. Until recently, they gave the best constraints on the LI violating parameter $\alpha$ (see~\cite{Will1992} for a review) with e.g. $|\alpha|\le10^{-6}$ obtained by comparing atomic clocks onboard GPS satellites with ground atomic clocks~\cite{Wolf1997}.

The three classical LI tests are the Michelson--Morley, Kennedy--Thorndike, and Ives--Stillwell experiments~\cite{robertson49}; they are second-order tests as the LI violating signal depends on $w^2/c^2$~\cite{mansouri77c}. With the advent of heavy-ion storage rings, Ives--Stillwell type experiments now give the best constraint on $\alpha$~\cite{Reinhardt2007,Botermann2014a}.  A limit of $|\alpha| \le 8.4\times10^{-8}$ was found using $^7$Li$^+$ ions prepared in a storage ring to~6.4\% and~3.0\% of the speed of light~\cite{Reinhardt2007}. The experiment described in~\cite{Botermann2014a} uses $^7$Li$^+$ ions confined at a velocity of~33.8\% of the speed of light. When neglecting higher order RMS parameters, the constraint on the LI violating parameter is $|\alpha| \lesssim 2.0\times10^{-8}$.

In this paper, we improve upon this best previous constraint on the LI violating parameter $\alpha$ by a factor of around two. Our test is based on four optical lattice clocks using Sr atoms, two located at LNE-SYRTE, Observatoire de Paris, France~\cite{0026-1394-53-4-1123,le2013experimental}, one at PTB, Braunschweig, Germany~\cite{1367-2630-16-7-073023,gre16}, and one at NPL, Teddington, UK~\cite{1742-6596-723-1-012019}. These clocks are connected by two fiber links, one running from SYRTE to PTB operated in June 2015~\cite{lisdat2015clock}, and one from SYRTE to NPL operated in June 2016. This paper exclusively uses the stability of the frequency comparisons between the clocks by looking for a periodic variation.

In a simplified setup, an optical clock comparison using a phase noise compensated fiber link can be described as a two-way frequency transfer between two observers A and B~\cite{Williams2008,Grosche2009,Grosche2014,Stefani2015,Gersl2015}. Observer~A emits an electromagnetic signal (e.g. an IR laser) with proper frequency $\nu_0$, received by observer~B at a proper frequency $\nu_1$, and partly reflected back to observer~A, where it is received with a proper frequency $\nu_2$. %The two-way Doppler signal $(\nu_2-\nu_0)$ is then locked to an ultra-stable reference -- thanks to a frequency shifter inserted along the route of the fiber -- such that
The ``redshift signal'' or de-syntonization is
%\begin{equation}
%	\frac{\nu_1}{\nu_0} = 1 + \frac{\nu_2-\nu_0}{2\nu_0} + \Delta \label{eq:delta} \ ,
%\end{equation}
\begin{equation}
	\Delta \label{eq:delta} = \frac{\nu_1 - \nu_0}{\nu_0} - \frac{\nu_2-\nu_0}{2\nu_0} \ .
\end{equation}
%where $\Delta$ is the ``redshift signal''~\cite{Will1992} or de-syntonization. 
The first term contains the relativistic redshift between the two observer locations as well as the first order Doppler shift, while the second term contains only the first order Doppler shift, realizing a well-known ``Doppler cancellation'' scheme~(see e.g. \cite{Will1992,Grosche2014}). The first term is measured by locally beating an optical clock with the electromagnetic signal at each end of the link, while the second term is fixed at a known value.

In the low-velocity limit the de-syntonization can be written as
\begin{equation}
	\Delta = \Delta_{cl} + \Delta_\alpha \label{eq:delta2} \ ,
\end{equation}
where $\Delta_{cl}$ contains the relativistic redshift due to the static part of the gravity potential as well as temporal variations. During the considered dates of clock comparisons, peak-to-peak fractional frequency variations up to $1.3\times10^{-17}$ between PTB and SYRTE, and up to $5\times10^{-18}$ between NPL and SYRTE are due to variations of the gravity potential induced by tides. Solid Earth and ocean tides are taken into account (see~\cite{Voigt2016}).

The LI violating term signal is:
\begin{equation}
	\Delta_\alpha = \alpha c^{-2} \left[ 2 \vw\cdot\left( \vv_A-\vv_B \right) + \left( v_A^2-v_B^2 \right) \right] + \Ol(c^{-3}) \ , \label{eq:delta_alpha}
\end{equation}
where $\vv_A$ and $\vv_B$ are respectively the velocities of clocks A and B in the non-rotating geocentric celestial reference system (GCRS). They are obtained by transforming the terrestrial coordinates of the clocks, considered as constant, with the SOFA routines~\cite{SOFA:2016-05-03}. $\ve{w}$ is the velocity of the Earth with respect to a preferred frame, taken as the rest frame of the cosmological microwave background (CMB). It is the sum of the Earth velocity with respect to the Solar System Barycenter (SSB), and the  SSB velocity with respect to the CMB. The celestial coordinates of the SSB velocity with respect to the CMB in galactic coordinates are 263.99\degree (longitude) and 48.26\degree (latitude)~\cite{Hinshaw2009}, which transformed to the GCRS give 11~h~11~m~36~s (right ascension) and $-6\degree$~54'~00'' (declination)~\cite{Krisher1990} with a norm of 369~km$\cdot$s$^{-1}$. In June 2015 and 2016 the norm of $\ve{w}$ was $w\simeq340$~km$\cdot$s$^{-1}$.

% time-variable gravity potential components induced by tides give approximately diurnal

The first term of the LI violation in equation~(\ref{eq:delta_alpha}) varies with a period of one sidereal day as the Earth rotates around its axis. It is therefore possible to bound the LI violating parameter $\alpha$ by looking for daily variations in the relative frequency difference $y$ between remote clocks, located at different longitudes (\emph{i.e.} different orientation of~$\ve{v}$) and/or different latitudes (\emph{i.e.} different norms of~$\ve{v}$). The second term of Eq.~(\ref{eq:delta_alpha}) is constant, and considering an upper bound of $2\times10^{-8}$ on the parameter $\alpha$~\cite{Botermann2014a}, is lower than $4\times10^{-20}$, which is significantly below the accuracy of the clocks. Therefore we do not take it into account in our model.

\begin{figure}[tb]
	\centering
	\includegraphics[width=\linewidth]{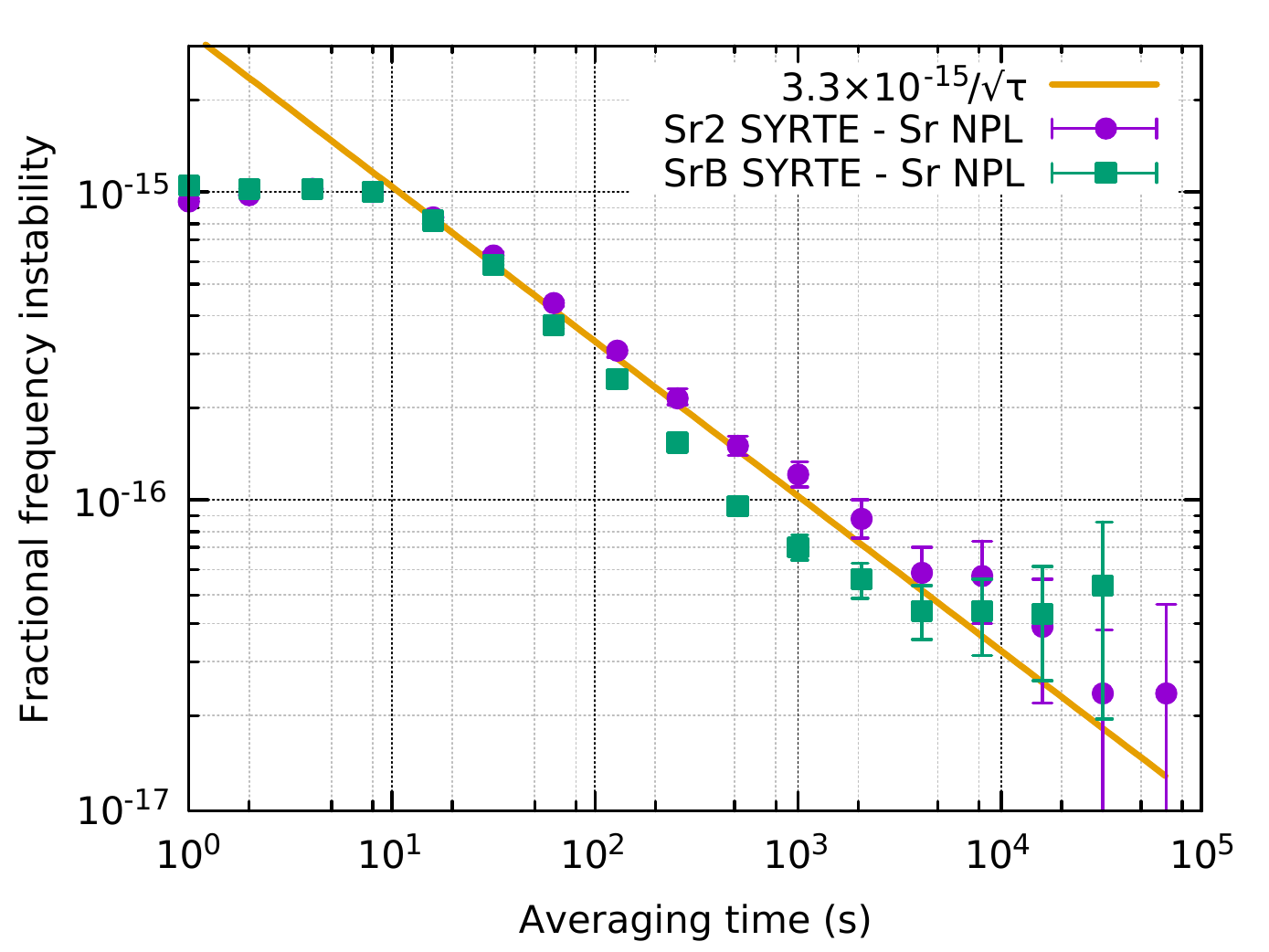}
	\caption{\label{fig:stability}Allan deviation of the fractional frequency difference between SYRTE's clocks Sr2 and SrB and NPL's Sr clock. After less than one day, the instability of the fractional frequency difference averages down to a few $10^{-17}$.
	This frequency instability is solely limited by the performances of the clocks, as the fiber link between SYRTE and NPL shows a fractional frequency instability of $1\times 10^{-18}$ at 1000~s.}
\end{figure}

We first analyze the result of the comparison between the clocks at SYRTE and NPL. Between June 10th and 15th 2016, we accumulated about 60~hours of clock comparison data between SYRTE's Sr2 and SrB lattice clocks, and NPL's Sr clock. These clocks are connected by a 812~km long cascaded optical fiber link using infra-red lasers operated at 1542~nm. The first span of 769~km connects NPL to Laboratoire de Physique des Lasers (LPL) in the north of Paris with the use of a repeater laser station at LPL\,\cite{Chiodo:2015}; the second link connects SYRTE to LPL\,\cite{Kroenjaeger2016, Lee2016}. The frequency ratio of the infra-red lasers and the Sr clock lasers at NPL and SYRTE are measured using optical frequency combs\,\cite{Johnson2015,nicolodi2014spectral}. The propagation noise in the fibers is actively compensated. At LPL, a beat note is generated with light from the two stabilized links and recorded using a GPS-disciplined ultra-stable quartz oscillator and a dead-time-free frequency counter with a similar approach to the set-up described in\,\cite{lisdat2015clock}. The frequency counters at NPL, LPL, and SYRTE are synchronized to UTC(NPL), GPS time, and UTC(OP) respectively with an accuracy well below 1~ms. Figure~\ref{fig:stability} shows the relative frequency instability of the comparison.

To search for a violation of LI in the clock comparisons, we consider three different data subsets: A: Sr2 data only; B: SrB data only; C: Sr2 and SrB data combined. The relative frequency difference $y_{\nplop}$ between the NPL Sr clock and the SYRTE Sr clock is corrected from the term $\Delta_{cl}$. The model used to fit the data contains two (for A and B subsets) or three parameters (for C subset):
%\begin{align}
%	y_{\nplop}(t) = \sum_{\rm i} \bar y^{\rm i}_{\nplop} \nonumber \\ 
%	+ 2 \alpha c^{-2} \vw\cdot\left[ \vv_{\op}(t)-\vv_{\npl}(t) \right] \ , \label{eq:model1}
%\end{align}
\begin{align}
	y_{\nplop}(t) = \bar y^{\rm i}_{\nplop} \nonumber \\ 
	+ 2 \alpha c^{-2} \vw\cdot\left[ \vv_{\op}(t)-\vv_{\npl}(t) \right] \ , \label{eq:model1}
\end{align}
where $\bar y^{\rm i}_{\nplop}$ allows for one or two fractional frequency offsets, depending on the chosen data subset: A: $\rm i=\{Sr2\}$; B: $\rm i=\{SrB\}$; C: $\rm i=\{Sr2,SrB\}$, and $\alpha$ is the LI violating parameter. All parameters are determined in the fitting procedure, along with correlations and uncertainties. The mean frequency offsets were removed from each comparison data subset as we are looking only for daily variations. The second line of Eq.~(\ref{eq:model1}) is the LI violation; it is very similar to a sinusoid: $Q_0\sin[2\pi(t-t_0)/T]$ where $T$ is one sidereal day, $Q_0=1.60\times10^{-10}$ for $\alpha=1$, and $t_0=57549.130$ (MJD).

\begin{figure}[thb]
	\centering
	\includegraphics[width=0.8\linewidth]{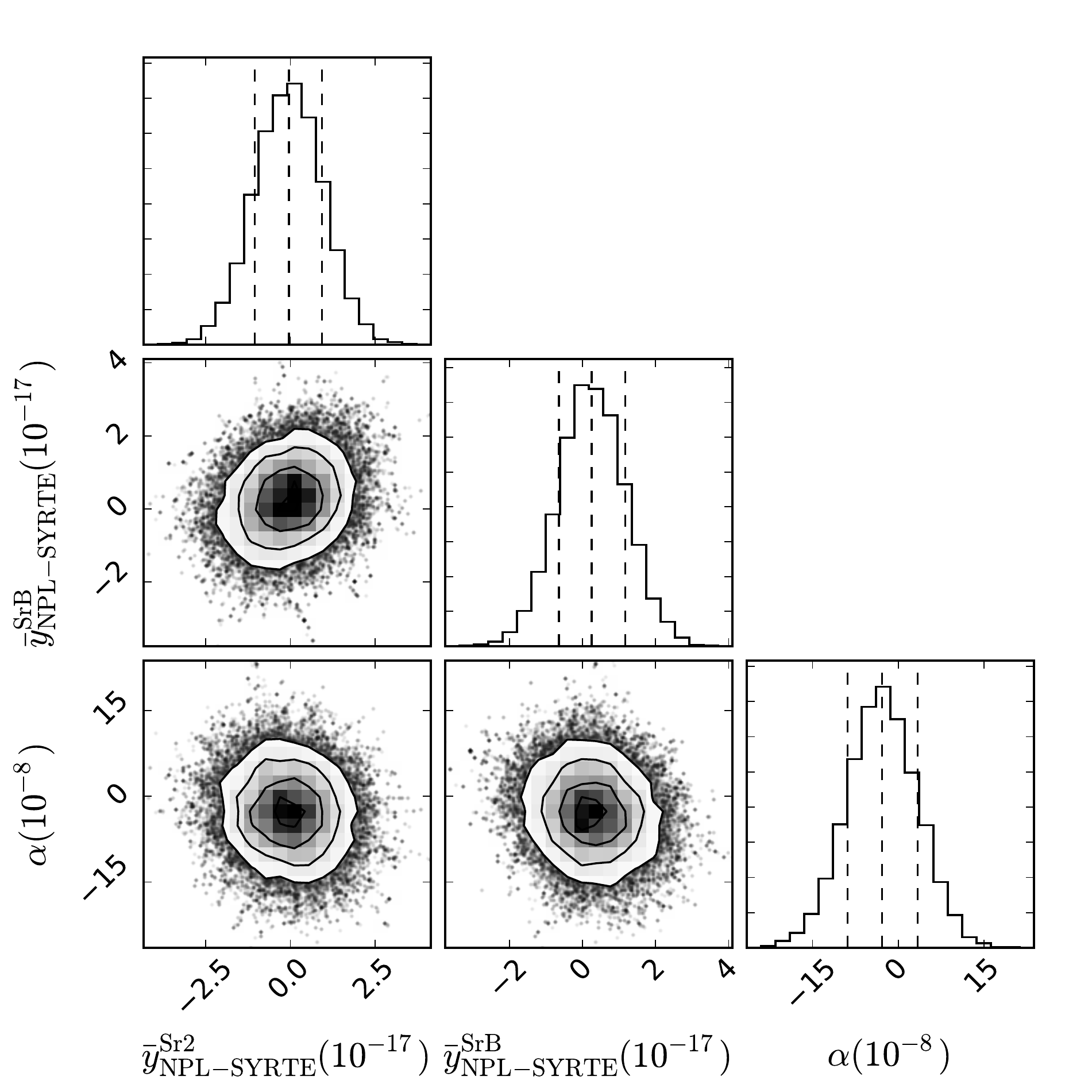}
	\caption{Correlations and histograms of the parameters of model~(\ref{eq:model1}) in fit C of table~\ref{tab:fit}, corresponding to the NPL-SYRTE comparison data from Sr2 and SrB combined, using the MCMC fitting method with $10^5$ points.\label{fig:npl}}
\end{figure}

For each data subset we used an affine invariant Markov Chain Monte Carlo ensemble sampler (MCMC) fitting method with $10^5$ points~\cite{Goodman2010}. As can be seen in Fig.~\ref{fig:stability}, the Allan deviation is flat up to around 10~s averaging time. Indeed, in the short term the laser probing the narrow transition is not yet locked to the atoms, therefore the flicker floor of the free running laser is visible. This temporal correlation is taken into account in the MCMC fit by using a non-diagonal covariance matrix. 
%Non-diagonal elements are calculated thanks to an exponential covariance function~\cite{rasmussen2006}, the parameters of which are fitted \emph{a priori} on the empirical covariance. 
For the combined data set~C the correlation between both data sets is also taken into account in the covariance matrix.

Fitting results are given in table~\ref{tab:fit} for the three cases~A to~C. The correlations between the parameters are of the order or below $0.2$. The best result is found when combining the two sets of data:
\begin{equation}
	\alpha_{\text{C}} = (-2.83\pm6.19)\times10^{-8}
	\label{eq:alphaC}
\end{equation}
Correlations and histograms of the parameters for data set~C are shown on figure~\ref{fig:npl}.

\begin{table}[b]
	\centering
	\begin{tabular}{|c||c|c|c|}
		\hline
		\rule[-0.2cm]{0cm}{0.6cm} & $\delta t^{\op}$ & $\alpha_T^{\op}$ & $\alpha$ \\
		& (hours) & ($10^{-16}$~K$^{-1}$) & ($10^{-8}$) \\\hline	
		A & -- & -- & $+3.81\pm8.41$\\\hline
		B & -- & -- & $-5.87\pm7.78$ \\\hline
		C & -- & -- & $-2.83\pm6.19$ \\\hline\hline
		D & $4.81\pm0.25$ & $1.76\pm0.12$ & $-0.38\pm1.06$ \\\hline
%		\bf combined & $\mathbf{4.67\pm0.24}$ & $\mathbf{1.77\pm0.12}$ & $\mathbf{-0.79\pm0.98}$ \\\hline	
	\end{tabular}
	\caption{\label{tab:fit}Fitting results using the MCMC fitting method with $10^5$ points. Fits A to C use the NPL-SYRTE comparison data with A: Sr2 data only; B: SrB data only; C: Sr2 and SrB data combined. Fit D uses the PTB-SYRTE comparison data.}
\end{table}

The PTB-SYRTE comparison took place between June 4th and 24th 2015. This comparison, involving SYRTE's Sr2 clock and PTB's stationary Sr clock, is reported in~\cite{lisdat2015clock}. We use in this paper the data of the second of the two campaigns reported in~\cite{lisdat2015clock}, representing around 150~hours of clock comparison data.

An analysis of the PTB-SYRTE comparison data with a model similar to equation~(\ref{eq:model1}) (replacing the NPL clock velocity with the PTB clock velocity, and $\rm i=\{Sr2\}$) results in a significant bias for the parameter $\alpha$, five times larger than the 1$\sigma$ uncertainty on $\alpha$. Indeed, the power spectral density distribution of the raw data shows a significant peak at a frequency of 1~day$^{-1}$, which is around the frequency of the LI violating signal. Although this signal could be interpreted as a violation of LI, a detailed analysis shows that this effect is probably due to temperature variations in the SYRTE clock laboratory. We analyzed two independent local clock comparisons: Sr against Yb$^+$ at PTB, and Sr against Hg at SYRTE, which are not affected by a LI violation, and we used a simple model of the effect of temperature on the relative frequency difference:
\begin{equation}
%	y_T (t;\alpha_T,\delta t) 
	y_{T,X} (t) 
	= \alpha_T \left[ T_{\text{X}}(t-\delta t) - \bar T_{\text{X}} \right] \ ,
	\label{eq:yT}
\end{equation}
where $T_\text{X}(t)$ is a function that interpolates the temperature at time $t$ at some location X, $\bar T_X$ is the mean of the temperature function $T_X(t-\delta t)$ evaluated for the comparison data times, $\alpha_T$ is a temperature coefficient and $\delta t$ a lag, both to be determined in the fitting procedure. A significant variation was found at 1~day$^{-1}$ frequency in the local SYRTE comparison, while the comparison at PTB did not show any significant variation at this frequency. 

Therefore, in addition to the LI violation, we included the effect of temperature in the SYRTE clock room, leading to the following model:
\begin{align}
	&y_{\ptbop}(t) =  \ySr \nonumber
	+ y_{T,\text{SYRTE}} (t) \\
	&+ 2 \alpha c^{-2} \vw\cdot\left[ \vv_{\op}(t)-\vv_{\ptb}(t) \right] \ ,
	\label{eq:model3}
\end{align}
where $\ySr$ allows for a fractional frequency offset, $y_{T,\text{SYRTE}}$ is the temperature effect model given in Eq.~(\ref{eq:yT}) and $\alpha$ is the LI violating parameter. The relative frequency difference $y_{\ptbop}$ between the PTB Sr clock and the SYRTE Sr clock is corrected from the term $\Delta_{cl}$. As for the model in Eq.~(\ref{eq:model1}), here the LI violating term is similar to a sinusoid with a period of one sidereal day, an amplitude $Q_0=3.54\times10^{-10}$ for $\alpha=1$, and $t_0=57177.421$ (MJD). As this is a non-linear model, we use the MCMC method with $10^5$ points for the fitting procedure. Note that $Q_0$ for the PTB--SYRTE link is more than twice the value for the NPL--SYRTE link such that it is more sensitive to a violation of LI.

\begin{figure}[t]
	\centering
	\includegraphics[width=\linewidth]{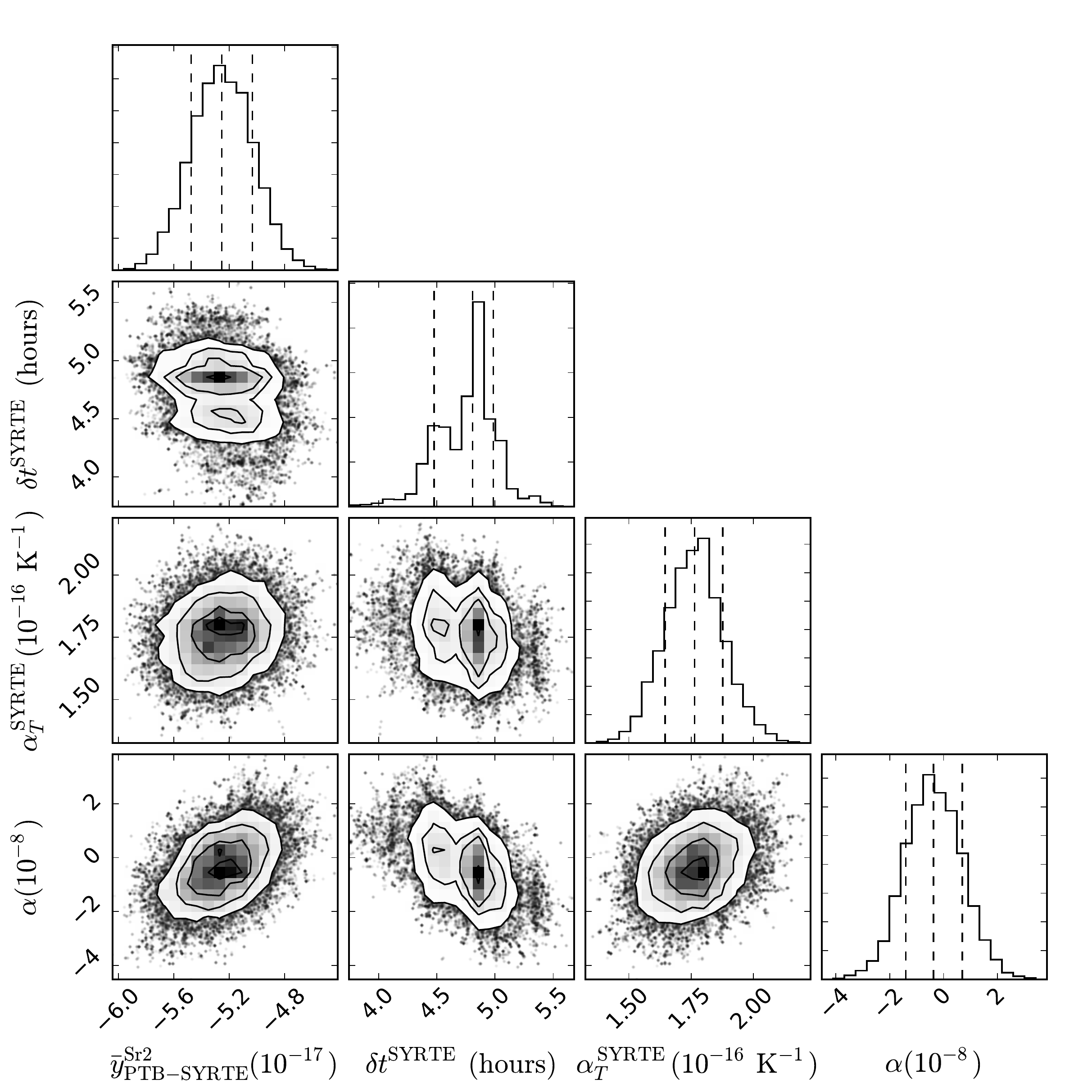}
	\caption{Correlations and histograms of the parameters of fit~D of table~\ref{tab:fit}, corresponding to the fit of the PTB-SYRTE comparison data, using the MCMC fitting method with $10^5$ points.}
	\label{fig:ptb}
\end{figure}

The detailed result of this analysis is given in table~\ref{tab:fit} -- line~D. It shows a significant effect of the temperature on the frequency comparison of the order of $10^{-16}$~K$^{-1}$, with a lag of around 4.8~hours. Correlations and distributions of parameters can be seen in figure~\ref{fig:ptb}. The lag $\delta t^{\op}$ is not well constrained and its distribution is non-Gaussian. However, this does not affect the Gaussianity of the other parameters. The correlations between the parameter $\alpha$ and the parameters $\ySr$, $\delta t^{\op}$ and $\alpha_T^{\op}$ are respectively $-0.48$, $0.44$ and 0.30. These correlations slightly degrade the uncertainty on the determination of $\alpha$. The bias of the parameter $|\alpha|$ is below the parameter uncertainty:
\begin{equation}
	\alpha_{\rm D} = (-0.38\pm1.06)\times10^{-8}
	\label{eq:alphaD}
\end{equation}
It is interesting to note that the bias found for the parameter $\alpha$ is $-0.93\times10^{-8}$ if tides are not taken into account.

In order to check further if the choice of this temperature is significant, we repeated the same analysis with several other temperature series: (i) the SYRTE exterior (ii) the PTB exterior (iii) the PTB clock room (iv) the PTB clock laser room (v) the PTB comb room and finally (vi) a simulated sinusoid with a 1~day period. For each temperature model we fit an amplitude and a lag, as in equation~(\ref{eq:yT}). None of them are able to explain the residual signal, leading for the series (i-v) to biases in the parameter $\alpha$ ranging from $2\sigma$ up to $5.5\sigma$, and for (vi) to a complete degeneracy of all parameters with no unique solution, which shows that the residual effect cannot be well represented by a simple sinusoidal model, as parameters are completely degenerated with parameters from the LI violating model. 

The effect of temperature on the PTB-SYRTE comparison data is not yet fully understood. Further comparisons will help improve our understanding of this, thereby allowing reduction of the bias and hence the uncertainty on the determination of the parameter $\alpha$. The existence of the temperature effect is however evident by the fact that it can be seen both on the distant PTB-SYRTE and the local SYRTE clock comparisons. A detailed analysis of the NPL-SYRTE comparisons did not show any significant systematic effects above the noise level. Consistently, a simulation of the temperature model~(\ref{eq:yT}) for the NPL-SYRTE comparisons, using the parameters determined in the PTB-SYRTE comparison, does not produce any signal above the noise level. This justifies the fact that the temperature model~(\ref{eq:yT}) was not used for the NPL-SYRTE comparisons.

A combination of the three data sets~A, B and~D has been evaluated but gives no improvement on the uncertainty of the determination of the LI violating parameter. This is due to the fact that the absolute value of the bias on~$\alpha$ obtained in combination~C ($2.83\times10^{-8}$) is larger that the uncertainty obtained with data set~D ($1.06\times10^{-8}$).

As noted in~\cite{Will1992}, one major limitation of the RMS framework is that it is purely kinematical, and our results cannot be simply mapped to dynamical frameworks. The constraints which can be derived from distant optical clock comparisons on dynamical frameworks such as the Standard Model Extension (see e.g.~\cite{Colladay1997,Colladay1998,kostelecky02,Tasson2014}) or Dark Matter models (see e.g.~\cite{Derevianko2014,Arvanitaki2015}) will be tackled in future publications.

In conclusion, by using clock comparisons between four optical clocks at NPL (UK), PTB (Germany) and SYRTE (France), linked by a leading-edge optical fiber network, we are able to put a more stringent bound on the LI violating parameter $\alpha$ of the RMS framework. With $1.1\times10^{-8}$, $\alpha$ is now by around a factor of two better constrained compared to the best previous determination of this parameter, which was obtained with accelerated ions, and by two orders of magnitude with respect to the best constraint previously obtained by comparing atomic clocks. Moreover, this bound is purely limited by technical noise sources on the clock systems, which will improve in future comparisons. Projecting the comparison of distant clocks with an instability of $10^{-16}/\sqrt{\tau}$ over several weeks, a reduction in uncertainty of more than one order of magnitude for $\alpha$ is within reach. This shows the significant potential for tests of fundamental physics with networks of optical clocks connected by optical fiber links.

\begin{acknowledgments}
We thank L. De Sarlo for providing the local SYRTE Sr/Hg local comparison data, J. Achkar for providing temperature data, and P. Wolf for useful discussions. We thank S. Koke and A. Koczwara for experimental assistance on the fiber link and Harald Schnatz for long-standing support. This work would not have been possible without the support of the GIP RENATER and G\'EANT. The authors are deeply grateful to E. Camisard  from GIP RENATER and G. Roberts  from G\'EANT. W.-K. Lee was supported partly by the Korea Research Institute of Standards and Science under the project ``Research on Time and Space Measurements'', Grant No. 16011007, and also partly by the R\&D Convergence Program of NST (National Research Council of Science and Technology) of Republic of Korea (Grant No. CAP-15-08-KRISS). This work was partly funded from the EC's Seventh Framework Programme (FP7 2007-2013) under Grant Agreement No. 605243 (GN3plus), and from the European Metrology Research Programme EMRP/EMPIR under SIB02 NEAT-FT, SIB55 ITOC, 15SIB05 OFTEN and 15SIB03 OC18. This project has received funding from the EMPIR programme co-financed by the Participating States and from the European Union's Horizon 2020 research and innovation programme. LNE-SYRTE and LPL acknowledge funding support from the Agence Nationale de la Recherche (ANR blanc LIOM 2011-BS04- 009-01, Labex First-TF ANR 10 LABX 48 01, Equipex REFIMEVE+ ANR-11-EQPX-0039), Action sp\'ecifique GRAM, Centre National d'\'Etudes Spatiales (CNES) and Conseil R\'egional \^Ile-de-France (DIM Nano-K). PTB acknowledges support from the German Research Foundation DFG within RTG~1729, CRC~1128 geo-Q, and CRC~1227 DQ-mat. W.B. and S.B. would like to acknowledge the support from the Marie Curie Initial Training Network FACT.
\end{acknowledgments}

\bibliography{RMS_fibre_prl}% Produces the bibliography via BibTeX.

\end{document}